\documentclass[
    a4paper,
    11pt,
    man,
    floatsintext
]{apa6}

\usepackage[english]{babel}
\usepackage[utf8]{inputenc}
\usepackage{epstopdf}
\usepackage{csquotes}
\usepackage[hidelinks]{hyperref}
\usepackage{kbordermatrix} 
\usepackage{amsmath} 
\usepackage{amssymb} 

\usepackage{natbib}

\title{Rank-deficiencies in a reduced information latent variable model}
\shorttitle{Rank deficiencies in a latent variable model}
\author{DL Oberski}
\affiliation{Utrecht University}


\abstract{
    Latent variable models are well-known to suffer from rank deficiencies, causing problems with convergence and stability. Such problems are compounded in the ``reduced-group split-ballot multitrait-multimethod model", which omits a set of moments from the estimation through a planned missing data design. This paper demonstrates the existence of rank deficiencies in this model and give the explicit null space. It also demonstrates that sample size and distance from the rank-deficient point interact in their effects on convergence, causing convergence to improve or worsen depending on both factors simultaneously. Furthermore, it notes that the latent variable correlations in the uncorrelated methods SB-MTMM model remain unaffected by the rank deficiency. I conclude that methodological experiments should be careful to manipulate both distance to known rank-deficiencies and sample size, and report all results, not only the apparently converged ones. Practitioners may consider that, even in the presence of nonconvergence or so-called "inadmissible" estimates, a subset of parameter estimates may still be consistent and stable.
}

\begin{document}

\maketitle

In a wide variety of fields, different data sources inform on the same phenomenon, the problem being to determine how these different sources should be combined, and how validly each measures the phenomenon of interest. For example, in official statistics, contradictory administrative registers and surveys may be available on citizens' employment contracts \citep{oberski_evaluating_2017,pankowska_reconciliation_2018}; in family sociology, reports from different family members may not always match up \citep{kenny_dyadic_2006}; and in medicine, a hospital may have data on patients' condition from electrocardiograms, echocardiograms, radiological examinations and individual laboratory measurements simultaneously  \citep{sammani_unravel:_2019}. In all such cases, latent variable models \citep{bartholomew_latent_2011} can  prove powerful tools to combine  different data sources measuring the same phenomenon in a principled manner \citep{hand_statistical_2018,oberski2018research}.  

A particularly useful approach is the ``multitrait-multimethod'' design, which was introduced by \citet{campbell_convergent_1959} to measure a single phenomenon (``trait'') using different data sources (``methods''), and to evaluate the sources' validity as measures of their underlying ``traits''. To analyze the resulting data, MTMM factor models were developed by \citet{browne1984decomposition,widaman_hierarchically_1985,cudeck_msultiplicative_1988,millsap_statistical_1995,wothke1995covariance}, and \citet{eid_multitrait-multimethod_2000}.   Extensions to nonlinear and nonnormal latent variable models were recently developed by \citet{oberski_evaluating_2017}. The advantage of MTMM models is that they recognize not only the common variance due to measurement of the same phenomenon, but also any common biases that arise from the use of a common data source. For example, survey answers may correlate due to ``acquiescence'' and social desirability bias \citep{lavrakas_extending_2019}, and electrocardiograms are susceptible to bias from manual annotators and placement of the electrodes \citep{zhu_fusing_2015}. MTMM models are designed to provide the researcher with an indication of the extent to which such biases are present and cause common correlation. At the same time, they determine implicit rules for the optimal guess regarding the true latent variable under study -- dispensing with the commonly used ad-hoc rules of data fusion. 

However, a disadvantage of latent variable models is that they are especially prone to problems of identification, nonconvergence, ``inadmissible'' estimates outside the acceptable range, and unstable estimates \citep{bartholomew_latent_2011}. MTMM models are especially well-known to suffer from such problems \citep{marsh_confirmatory_1989,brannick_estimation_1990,kenny_analysis_1992,marsh_overcoming_1992,bagozzi_assessing_1993,revilla_split-ballot_2013}. Solutions to this problem have also been suggested. For example, \citet{marsh_overcoming_1992} suggested omitting a model for all method factors while acconting for the resulting correlations; \citet{eid_multitrait-multimethod_2000} suggested omitting one method factor from the model; and \citet{castro-schilo_augmenting_2016} suggested use of an additional, completely independent, data source for each trait; and 
\citet{saris_comparing_2019} suggested leveraging information from multiple groups (countries) to aid estimation. Each of these solutions has its relative merits and disadvantages; specifically, each requires either the abandonment of method effects as target parameters, or additional information that may only sometimes be available. For these reasons, the final solution to estimation problems with MTMM is still under expert discussion.

In spite of the existence of suggested solutions--and regardless of their relative merits--relatively little is known about the \emph{cause} of estimation problems in MTMM models: rank deficiencies in the model's information matrix. In this paper, we will investigate that cause in analytical detail, and indicate exactly how it operates to generate nonconvergence and parameter instability (``inadmissible'' estimates, bias in constrained estimation). To do this, we will use a particularly problematic version of the MTMM model as case study: the ``reduced-group split-ballot multitrait-multimethod'' model \citep{saris_new_2004}. This model is problematic because it involves a planned missing data design, and therefore estimation provides even more limited information about the parameters than is usually the case in MTMM. 

\medskip
The following section first defines the factor or structural equation (SEM) model framework that is commonly used in MTMM. It also explains how nonconvergence and ``inadmissible'' estimates can occur, even when the model is correctly specified. The relationship of rank deficiency with identification is explained, and some intuition regarding rank deficiencies of zero probability measure are given. The subsequent section defines the reduced-group correlated trait-uncorrelated method SB-MTMM model used by \citet{saris_design_2014,revilla_split-ballot_2013}, and \citet{saris_comparing_2019}, and derives the rank deficiency that occurs in this model using a computer algebra system. We then perform two Monte Carlo experiments that demonstrate the consequence of such rank deficiencies in simple factor models, as well as the CTUM RG-SB-MTMM model. Finally, the conclusion reflects on the role different types of parameters play in the estimation, the paradoxical role of sample size,  the need to report nonconverged results in simulation studies, and, finally, the connection to modern unsupervised machine learning versions of LVM's and regularization. It is hoped the methods presented in this paper can form the basis for routine evaluation of latent variable models, and can provide insight in the potential solutions to their woes.

\section{Background}

Let $\textbf{y} \in \mathbb{R}^q$ be an $q$-vector of observable variables. For simplicity, we will assume all variables are centered.
The basic confirmatory factor analysis (CFA) model is then
\begin{equation}
    \textbf{y} = \Lambda \boldsymbol{\eta} + \boldsymbol{\epsilon},
\end{equation}
where the ``common factors'' $\boldsymbol{\eta} \in \mathbb{R}^{q^*}$ and ``residuals'' $\boldsymbol{\epsilon} \in \mathbb{R}^q$ are unobserved vectors of latent variables. Generally, there are fewer factors than observed variables, $q^*<q$, and we assume the factors and residuals are uncorrelated, $\mathbb{E}(\boldsymbol{\eta} \boldsymbol{\epsilon})  = \mathbf{0}$, and both latent variable vectors have constant variance matrices, say, $\text{Var}(\boldsymbol{\eta}) = \boldsymbol{\Phi}$, and $\text{Var}(\boldsymbol{\epsilon}) = \boldsymbol{\Psi}$. 

The parameters of interest of the factor model are the loading matrix $\Lambda$, the factor variance matrix $\boldsymbol{\Phi}$, and the residual variance matrix $\boldsymbol{\Psi}$. Generally, these matrices are sparsely parameterized, so that only certain elements are free parameters of the model to be estimated. We collect these free parameters into a single parameter vector, $\boldsymbol{\theta} = (\boldsymbol{\lambda}^T, \boldsymbol{\phi}^T, \boldsymbol{\psi}^T)^T$, say, of length $p$.  Under the above assumptions of linearity and homoskedasticity, the implied variance of the observed variable vector is then
\begin{equation}\label{eq:sigma}
    \text{Var}(\mathbf{y} | \boldsymbol{\theta}) = \Sigma(\boldsymbol{\theta}) = \Lambda \boldsymbol{\Phi} \Lambda^T + \boldsymbol{\Psi}.
\end{equation}
The parameters of interest $\boldsymbol{\theta}$ can be estimated in a sample through maximum-likelihood with $\Sigma(\boldsymbol{\theta})$ as the covariance matrix, or, equivalently, by minimizing the weighted least squares loss,
\begin{equation}
\hat{\boldsymbol{\theta}}_n = 
        \arg\min_{\boldsymbol{\theta} \in \Theta} F,
\label{eq:objective}    
\end{equation}
where the loss function $F$ is the weighted sum of squared residuals,
\begin{align}
    \label{eq:wls}
        F(\boldsymbol{\theta}) =
            \left[\mathbf{s}_n - \boldsymbol{\sigma}(\theta)\right]^T
            \mathbf{V}
            \left[\mathbf{s}_n - \boldsymbol{\sigma}(\theta)\right],
\end{align}
and $\mathbf{s}_n = \text{vech}(\mathbf{S}_n)$ is the half-vectorized observed covariance matrix obtained from an i.i.d. sample of size $n$, with $\boldsymbol{\sigma}(\boldsymbol{\theta}) = \text{vech}[\boldsymbol{\Sigma}(\boldsymbol{\theta})]$, of length $p^*$.  Throughout, we will assume that both the observed and the population covariance matrices are positive-definite.  (This assumption is violated, for example, in the case of high-dimensional data with $q>n$.)

Setting $$\mathbf{V} := 2^{-1} \mathbf{D}^T (\hat{\boldsymbol{\Sigma}}^{-1} \otimes \hat{\boldsymbol{\Sigma}}^{-1}) \mathbf{D},$$ where $\mathbf{D}$ is the "duplication matrix", yields normal-theory maximum likelihood estimation \citep{neudecker_linear_1991}. Here $\hat{\boldsymbol{\Sigma}}$ is a consistent estimate of the population covariance matrix $\text{Var}(\mathbf{y})$. Some procedures set $\hat{\boldsymbol{\Sigma}} := \boldsymbol{\Sigma}(\hat{\boldsymbol{\theta}})$ iteratively during estimation, while others use $\hat{\boldsymbol{\Sigma}} := \mathbf{S}_n$. More generally, we will assume $\mathbf{V}$ to be any positive-definite matrix of estimation weights. 

\subsection{Estimation, nonconvergence, and ``inadmissible'' estimates}

Sample parameter estimates $\hat{\boldsymbol{\theta}}_n$ are generally found through an optimization procedure with objective given in Equation \ref{eq:objective}. The gradient of $F$ with respect to the parameters plays a key role in convergence of any such procedure. For example, in gradient descent optimization, the updating step at iteration $t+1$ is 
\begin{equation}
    \hat{\boldsymbol{\theta}}_{t+1} \leftarrow \hat{\boldsymbol{\theta}}_t - \boldsymbol{A}_t \cdot \boldsymbol{g}(\hat{\boldsymbol{\theta}}_{t}),
\end{equation}
where $\boldsymbol{g}(\hat{\boldsymbol{\theta}}_{t})$ is the gradient vector at iteration $t$, $\boldsymbol{g} = \dot{F}(\boldsymbol{\theta})$, and $\boldsymbol{A}_t$ a  "learning rate" matrix. Common choices for the learning rate in SEM software are the observed information at step $t$, i.e. $\boldsymbol{A}_t := \ddot{F}(\boldsymbol{\theta})^{-1}$, which gives Newton-Raphson optimization; the expected information, $\boldsymbol{A}_t := \left(\boldsymbol{\Delta}^T \mathbf{V} \boldsymbol{\Delta}\right)^{-1}$, which gives Fisher scoring; or an approximation to the inverse observed Hessian used in quasi-Newton methods such as BFGS. In the machine learning literature, other gradient-baed methods have been developed, with a baseline choice being $\boldsymbol{A}_t:=\gamma$, a scalar constant learning rate 
\citep[for a short overview of various optimization methods, see][Ch. 8]{goodfellow_deep_2016}. 

Convergence of these optimization algorithms is achieved when the gradient vector equals zero. However, when the gradients are linearly dependent, convergence will never be achieved, since the norm of the gradient will not decrease along the line $\mathbf{n} \cdot \boldsymbol{g} = \mathbf{0}$ for some $\mathbf{n} \neq \mathbf{0}$, by the definition of linear dependence. Here, $\mathbf{n}$ is a non-trivial (nonzero) basis for the ``nullspace'' of the gradient.  Note that the same holds for non-gradient based optimization methods; for example,  proof of convergence of the Nelder-Mead algorithm requires the absence of linear dependencies in the gradient as well \citep[e.g.]{lagarias_convergence_1998}.
For SEM, this gradient is 
\begin{equation}
     \boldsymbol{g} = \dot{F}(\boldsymbol{\theta})  = \boldsymbol{\Delta}^T \mathbf{V} \left[\mathbf{s} - \boldsymbol{\sigma}(\boldsymbol{\theta})\right],
\end{equation}
where $\boldsymbol{\Delta}$ is the Jacobian of the implied (co)variances with respect to the parameters, $\boldsymbol{\Delta} = \dot{\boldsymbol{\sigma}}(\boldsymbol{\theta})$. This Jacobian has $p^*$ (the number of unique variances and covariances) rows and $p$ (number of parameters) columns; note that the degrees of freedom of the model equals $p^* - p$. Drawing a parallel with linear regression, the matrix  $\boldsymbol{\Delta}$ can be thought of as a design matrix in the linearized mapping of parameters into (co)variances \citep{savalei_understanding_2014}.
Since  $\boldsymbol{S}$, $\boldsymbol{\Sigma}$, and $\mathbf{V}$ are positive-definite by assumption, linear dependence in the gradients can only occur as a consequence of a rank deficiency in the Jacobian $\boldsymbol{\Delta}$. Thus, nonconvergence is a direct consequence of rank deficiency of $\boldsymbol{\Delta}$, i.e. when the column rank $\text{rk}(\boldsymbol{\Delta}) < p$.  

So-called ``inadmissible'' estimates also result from rank deficiencies in the Jacobian $\boldsymbol{\Delta}$. ``Inadmissible'' estimates -- such as negative estimates for variance parameters or non-positive definite latent variable corvariance matrices -- are possible because the optimization space is usually taken as $\mathbb{R}^p$, which includes ``inadmissible'' subsets. In fact, for many standard CFA models, \emph{most} of the optimization space is ``inadmissible'' \citep[see][for the closely related concept of``complexity'' of inequality-constrained Bayesian models as the admissible probability mass]{mulder_equality_2010}.
For this reason, as the variance of the estimates increases, so does the the probability of so-called ``inadmissible'' sample estimates of the parameter vector. This holds regardless of whether the ``population parameter vector'' -- the solution to which Equation \ref{eq:objective} convergences as $n$ grows without bound -- is not itself ``inadmissible''. A separate case is misspecification of the model, which is the most widely-recognized cause of  ``inadmissible'' population parameter vectors  \citep{chen_improper_2001}. Here we will ignore such cases, and assume the model is correctly specified. We will see that even in this idealized situation, severe problems with estimation can occur when there are rank-deficient points in the Jacobian.

Under the assumption of a correctly specified model, variance of SEM estimates $\hat{\boldsymbol{\theta}}_n$ is obtained, through standard likelihood theory, as the inverse Fisher information, 
\begin{equation}\label{eq:variance}
    \text{Asy.}\text{Var}(\hat{\boldsymbol{\theta}}_n) = \left(\boldsymbol{\Delta}^T \mathbf{V} \boldsymbol{\Delta}\right)^{-1}.
\end{equation}
Relevant theory and extensions to more general settings can be found in \citep{satorra_alternative_1989}. Again, since $\mathbf{V}$ is positive-definite by assumption, the inverse in Equation \ref{eq:variance} will grow without bound as the Jacobian $\boldsymbol{\Delta}$ approaches singularity, causing bad performance in terms of MSE, as well as ``inadmissible'' estimates.

Some authors \citep[e.g.][]{rindskopf_parameterizing_1983} have suggested solving the problem of ``inadmissible'' estimates by excluding them from the optimization space. The same idea appears the norm in the literature on Bayesian SEM \citep{lee_structural_2007,muthen_bayesian_2012,merkle_blavaan:_2018}. However, others have pointed out that constrained estimation to prevent ``inadmissible'' estimates creates bias in the parameter estimates \citep[e.g.][]{chen_improper_2001}. The estimates that would otherwise have been ``inadmissible'' will ``pile up'' along the constraint boundary, neglecting to cancel out sample estimates further to the opposite side of the true parameter value. In other words, 
the underlying problem is not a computational one, and ``solving'' it with a computational trick such as constrained estimation will simply transfer the problem to a different part of the overall procedure, like pushing on an air mattress to deflate it while forgetting to open the valve. Due to this ``air matress'' principle, we will refer mostly to the problem of ``inadmissible'' estimates, with the understanding that the reader who prefers constrained estimation may mentally substitute this for the problem of bias.

In discussions of rank-deficiencies of parametric models, the information matrix inverted in Equation \ref{eq:variance} is taken as a point of departure \citep[e.g.][]{wald1950note,bekker_identification_2001}. In the case of linear and homoskedastic (SEM) models with full-rank estimation weight matrix $\mathbf{V}$, rank-deficiency of the information matrix is equivalent to rank-deficiency of the Jacobian of the sufficient statistics, $\dot{\boldsymbol{\sigma}}(\boldsymbol{\theta}) = \boldsymbol{\Delta}$. Because of this equivalence, and because the Jacobian has a much simpler form than the information matrix, we will focus here on the Jacobian.







\subsection{When is the Jacobian rank deficient?}

We have seen that the Jacobian $\boldsymbol{\Delta}$ plays a central role in generating nonconvergence and ``inadmissible'' estimates (or bias), specifically when this matrix is rank-deficient.  But when do such deficiencies occur? Here, we will distinguish two cases: underidentification and singular points. 

Underidentification is the most well-known cause of rank deficiency of $\boldsymbol{\Delta}$. For example, when the degrees of freedom are negative, $p^* < p$ it is obvious that the $p^* \times p$ matrix $\boldsymbol{\Delta}$ will not have full column rank. The same occurs whenever the implied (co)variances of the model, and therefore the likelihood for which these are sufficient statistics, are equal for two different sets of parameter values, i.e. $\boldsymbol{\sigma}(\boldsymbol{\theta}) = \boldsymbol{\sigma}(\boldsymbol{\theta}^\prime)$ but $\boldsymbol{\theta} \neq \boldsymbol{\theta}^\prime$ \citep[e.g.][]{wald1950note,bekker_identification_2001}. The absence of this problem is referred to as ``local'' identification in the literature when the condition needs to hold only for an open neighborhood of the parameter vector, rather than for every point of the parameter space. Note that (local) underidentification is not related to the sample at hand; it is a property of the population model. 

While local underidentification causes a rank-deficient Jacobian, the converse is \emph{not} true \citep{shapiro_investigation_1983}: when the Jacobian is rank deficient, it is not necessarily the case that, for all samples, $\boldsymbol{\sigma}(\boldsymbol{\theta}) = \boldsymbol{\sigma}(\boldsymbol{\theta}^\prime)$ (the model is underidentified). Generally, this can occur because the reverse implication holds only when the loss function is twice differentiable and the Jacobian has constant rank in a neighborhood around the parameter vector. When there are single parameter points at which the rank becomes deficient, this last assumption is violated. A well known example is the two-factor model with two indicators for each factor, which has a rank deficiency when the correlation among factors is zero. Since the probability of finding such a rank-deficient point  \emph{exactly} equals zero, the model is said in the  literature to be locally identified ``almost everywhere'', i.e. everywhere except in points with probability measure zero \citep{shapiro_identifiability_1985}. 

\begin{figure}[tb]
    \centering
    \includegraphics[width=\textwidth]{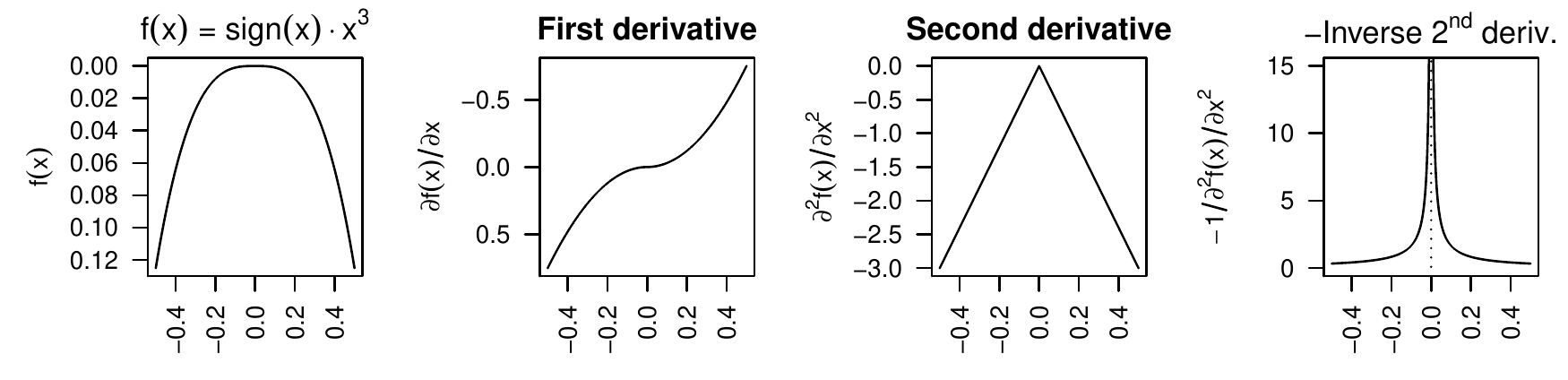}
    \caption{Example of a rank-deficient function, $f(x) = \text{sign}(x) \cdot x^3$. The function has a unique maximum at $x=0$, but at this point its second derivative is zero and changes discontinuously. The rightmost panel shows the negative inverse second derivative, which plays the role of the variance in likelihood theory, has an asymptote at this point.}
    \label{fig:fx}
\end{figure}

To illustrate intuitively how it is possible to have a rank-deficient information matrix (second derivative) at a singular point but no identification problem, consider Figure \ref{fig:fx}. The Figure illustrates the function $f(x) = \text{sign}(x) \cdot x^3$. In this example, $f(x)$ plays the role of the (log)``likelihood'', $x$ is the parameter, the first and second derivatives (second and third panels) are the gradient and Hessian, and the fourth panel plots the inverse second derivative, which plays the role of the variance of the parameter estimate. It can be seen in Figure \ref{fig:fx} that, while the second derivative is rank-deficient at the point $x=0$, which also happens to be the maximum, this maximum is still uniquely identifiable. However, as the point is approached, the variance of any finite-sample estimate of the parameter will grow without bound (asymptote in fourth panel). 

Figure \ref{fig:fx} shows that estimation problems do not only occur at the ``almost surely'' impossible point $x=0$. Points \emph{close} to this ``point of deficiency'' (p.o.d.) will also generate extremely high variance of the sample estimates, causing inadmissibility and instability, and nearly singular gradients, generating nonconvergence. Therefore, although the probability of solutions to the objective being at the p.o.d. \emph{exactly} is zero, the p.o.d. creates a zone of near-deficiency that will cause serious estimation problems in practice. Others have noted this issue as well; because of the dependency of practical problems on data, \citet{aldrich_how_2002} suggested to abandon the idea of ``identification'' and suggested replacing it with a concept of the informativeness of data. \citet[ch. 8]{goodfellow_deep_2016} discuss how the neural network community shifted its focus from the investigation of rank deficiencies to evaluating whether solutions that result are acceptable in terms of the cost function. In short, the econometric concept of ``almost sure'' stability may not be as comforting as it sounds. At the same time, the existence of rank deficiencies need not always generate serious problems with the estimation.

Finding rank-deficient points in the Jacobian is a challenging task. For factor analysis, and specifically multitrait-multimethod models, existing literature has developed several analytical results \citep[see][and references therein]{shapiro_identifiability_1985,kenny_analysis_1992,grayson_identification_1994}. An alternative approach is to employ computer algebra systems to investigate rank deficiency \citep{bekker_identification_2015}. This is what the following section will do for the reduced-group split-ballot multitrait multimethod (SBMTMM) model.

\section{Rank-deficiencies in reduced-group split-ballot multitrait-multimethod models}

The split-ballot MTMM design \citep{saris_new_2004} is a two-group randomized design in which different elements of $\mathbf{y}$ are observed for two groups of subjects. This yields a planned missing data design. In the reduced-group SBMTMM design, not all possible combinations of methods are observed. For example, in the European Social Survey (ESS), group 1 receives a questionnaire with versions 1 and 2 of a questionnaire (method 1 and 2), whereas group 2 receives a questionnaire with versions 1 and 3 (methods 1 and 3). Therefore, all covariances between measures obtained with methods 2 and 3 are completely missing. \citet{revilla_split-ballot_2013} noted that this particular design yields serious problems with nonconvergence, instability, and inadmissible estimates. 

Here, we will follow the ESS design discussed by \citet{revilla_split-ballot_2013}: three traits, three methods, and two groups in which methods 1 and 2 (group 1) and methods 1 and 3 (group 2) have been used. Let $\mathbf{y}_g$ be the vector of observed variables for group $g$, and let $y_{tm}$ indicate the measure of the $t$-th trait obtained with the $m$-th method. The observed variable vectors are then $\mathbf{y}_1 = (y_{11}, y_{12}, y_{21}, y_{22}, y_{31}, y_{32})^T$ (group 1) and $\mathbf{y}_2 = (y_{11}, y_{13}, y_{21}, y_{23}, y_{31}, y_{33})^T$ (group 2). Note that neither group contains both methods 2 and 3. As above, the CFA model per group is
\begin{equation}
    \label{eq:cfa-group}
    \boldsymbol{\Sigma}_g(\boldsymbol{\theta}) = \Lambda_g \Phi_g \Lambda_g^T + \Psi_g,
\end{equation}
where subscripts $g$ indicate group specific vectors. To facilitate further analysis, we redefine the vector of (co)variances as $\boldsymbol{\sigma}(\boldsymbol{\theta}) := (\boldsymbol{\sigma}_1(\boldsymbol{\theta})^T, \boldsymbol{\sigma}_2(\boldsymbol{\theta})^T)^T$, thus deleting the unobserved moments, and will do the same for observed moments $\mathbf{s}$.

As in \citet{revilla_split-ballot_2013}, we then specify the ``correlated-trait uncorrelated-method'' (CTUM) model with equal method loadings,
\begin{align}\label{eq:lambda-1}
\Lambda_1 =    \left(
\begin{array}{ccccc}
 \lambda_{11} & 0 & 0 & 1 & 0 \\
 \lambda_{12} & 0 & 0 & 0 & 1 \\
 0 & \lambda_{21} & 0 & 1 & 0 \\
 0 & \lambda_{22} & 0 & 0 & 1 \\
 0 & 0 & \lambda_{31} & 1 & 0 \\
 0 & 0 & \lambda_{32} & 0 & 1 \\
\end{array} 
\right)&\;& 
    \Lambda_2 = \left(
\begin{array}{ccccc}
 \lambda_{11} & 0 & 0 & 1 & 0 \\
 \lambda_{13} & 0 & 0 & 0 & 1 \\
 0 & \lambda_{21} & 0 & 1 & 0 \\
 0 & \lambda_{23} & 0 & 0 & 1 \\
 0 & 0 & \lambda_{31} & 1 & 0 \\
 0 & 0 & \lambda_{33} & 0 & 1 \\
\end{array}
\right),
\end{align}
standardized trait and method factors, with uncorrelated method factors,
\begin{align}
    \Phi_1 =  \kbordermatrix{
     &   \eta_1 &\eta_2 & \eta_3 & \xi_1& \xi_2\\
 \eta_1  &  1 & \rho_{12} & \rho_{13}  & 0& 0\\
 \eta_2 &  \rho_{12} & 1 & \rho_{23} & 0& 0\\
 \eta_3  &  \rho_{13} & \rho_{23} &1 & 0& 0\\
  \xi_1 &  0& 0& 0 & 1& 0\\
 \xi_2  &  0& 0& 0 & 0& 1\\
}, &\;& 
 \Phi_2 =  \kbordermatrix{
     &   \eta_1 &\eta_2 & \eta_3 & \xi_1& \xi_3\\
 \eta_1  &  1 & \rho_{12} & \rho_{13}  & 0& 0\\
 \eta_2 &  \rho_{12} & 1 & \rho_{23} & 0& 0\\
 \eta_3  &  \rho_{13} & \rho_{23} &1 & 0& 0\\
  \xi_1 &  0& 0& 0 & 1& 0\\
 \xi_3  &  0& 0& 0 & 0& 1\\
},
\end{align} 
and error variance matrices
\begin{align}
    \Psi_1 = \text{diag}(\psi_{1},\psi_{ 2},\psi_{3},\psi_{ 4},\psi_{5},\psi_{6}), &\;& 
    \Psi_2 = \text{diag}(\psi_{1},\psi_{7},\psi_{3},\psi_{8},\psi_{5},\psi_{9}).
\end{align}

The Jacobian of this model, $\boldsymbol{\Delta}_{\text{\sc{sbmtmm}}}$, is given in Equation \ref{eq:delta-ctum} in the Appendix.
When applying a computer algebra system such as Mathematica \citep{Mathematica} to this problem, we obtain a non-trivial nullspace \emph{only} in the following conditions:
\begin{enumerate}
    \item All loadings are equal, $\lambda_{tm} = \lambda$ \emph{and};
    \item All correlations are equal, $\rho_{t t^\prime} = \rho$.
\end{enumerate}
In this case, a basis for the nullspace is 
\begin{equation}\label{eq:nullspace}
\text{Null}(\boldsymbol{\Delta}_{\text{\sc{sbmtmm}}}) = \kbordermatrix{
       & \lambda_{11} & \ldots & \lambda_{12} & \ldots & \psi_{1} & \psi_{2} & \ldots &\rho_{12} & \rho_{13} & \rho_{23} & \phi_{4} & \phi_{5} & \phi_{6} \\
& \frac{1}{2 \lambda \rho} & \ldots & -\frac{1}{2 \lambda \rho} & \ldots & +\frac{\rho-1}{\rho} & -\frac{\rho-1}{\rho} & \ldots & 0 & 0 & 0 & -1 & 1 & 1 \\
}
\end{equation}
Crucially, $\text{Null}(\boldsymbol{\Delta}_{\text{\sc{sbmtmm}}})$ has zeroes in the three places that correspond to the three correlation parameters $\rho_{12}$, $\rho_{13}$, and $\rho_{23}$. The nullspace is orthogonal to these parameters, which are not involved in the dependency. We will see in the experiments that this means that models that appear to lack convergence, will actually converge for these three parameters, and estimates of these parameters will be stable in spite of high-variance (often inadmissible) estimates for the others.

The result that the Jacobian is deficient \emph{only} under the above two conditions may seem somewhat surprising given the literature on MTMM. 

First,  \citet{wald1950note} and \citet{kenny_analysis_1992} suggested that rank-deficiency of $\Lambda$ would generate an ``underidentified'' model. \citet{grayson_identification_1994} showed that this is correct, for the CTM and CTCM models, which introduces correlation parameters among methods (CTCM) and, additionally, between methods and traits (CTM). Second, \citet[p. 130]{grayson_identification_1994} also suggested conditions  under which the CTUM under consideration here would be identified. These conditions are met under conditions 1 and 2 above. However, these authors did not consider the \emph{reduced-group} split-ballot model; in this, present, model, the missing heteromethod moments generate the rank deficiency above. This fact explains observations by \citet{revilla_split-ballot_2013} that the observed nonconvergence and stability issues with reduced-group SBMTMM disappear when a third group including these heteromethod moments is included in the analysis. Finally, \citet{saris_new_2004} remarked that rank-deficiency of the reduced-group SBMTMM model occurs whenever condition 2 above is fulfilled. Our analysis indicates that this is not sufficient to generate a rank deficiency, but both conditions 1 and 2 must be fulfilled. At the same time, as noted by \citet{revilla_split-ballot_2013}, problems are indeed observed primarily when condition 2 is approached. The following sections bear out this observation.

\section{Experiments}

\subsection{Shapiro's example: rank deficiency versus identification}

To illustrate the problem of rank deficiencies in a simpler case than reduced-group split-ballot multitrait-multimethod models, we will first discuss a simple classical example discussed by \citet{shapiro_investigation_1983}. 

\citet{shapiro_investigation_1983} stressed that rank-deficiency of the information matrix need not imply a non-identified model. Figure \ref{fig:fx} demonstrated how this is possible: a function may well have a unique maximum even though it does not have a full-rank second derivative at every point. In this case, as noted by \citet{shapiro_investigation_1983}, the regularity conditions suggested by \citet{wald1950note} are violated, as the second derivative is not constant within any neighborhood around the maximum, but changes abruptly when moving away from this point (third panel in Figure \ref{fig:fx}). For this reason, the rank condition on the information matrix is only indicative of a true identification problem when this regularity condition is met.

\citet{shapiro_investigation_1983} illustrated this point with a three-indicator confirmatory factor model reparameterized as  $\boldsymbol{y} = \Lambda \boldsymbol{\eta}$ with 
\begin{align}
    \Lambda = \kbordermatrix{
      &  \eta & \epsilon_1 & \epsilon_2 & \epsilon_3 \\
       y_1 &  \lambda_1 &  \psi_1 & 0 & 0\\
       y_2 &  \lambda_2 & 0 & \psi_2 & 0 \\
         y_3 &\lambda_3 & 0 & 0 & \psi_3 \\
    }, & &
    \Phi = \text{Var}(\boldsymbol{\eta}) = \mathbf{I}_4.
\end{align}
All implied variances and covariances can then be written $\sigma_{jj^\prime}(\boldsymbol{\theta}) = \lambda_j \lambda_{j^\prime} + \delta_{j j^\prime} \psi_j^2$, where $\delta_{j j^\prime}$ is an indicator function that equals 1 if $j=j^\prime$ and 0 otherwise. This parameterization ensures that the implied error variances, $\psi_j^2$, are positive, even though all parameters are reals.

The Jacobian of this model is
\begin{equation}
    \Delta_{\text{Shapiro}} = \kbordermatrix{
        &\lambda_1 & \lambda_2 &\lambda_3 & \psi_1 & \psi_2& \psi_3\\
\sigma_{11} & 2 \lambda_1 & 0& 0& 2 \psi_1 & 0& 0\\
\sigma_{21} & \lambda_2  & \lambda_1 & 0& 0& 0& 0\\
\sigma_{31} & \lambda_3  & 0 & \lambda_1 & 0& 0& 0\\
\sigma_{22} & 0 & 2 \lambda_2 & 0& 0 & 2 \psi_2& 0\\
\sigma_{32} & 0  & \lambda_3 & \lambda_2 & 0& 0& 0\\
\sigma_{33} & 0 & 0 & 2 \lambda_3 & 0 & 0& 2 \psi_3\\
},
\end{equation}
so that simply setting any $\psi_j = 0$ will lead to a rank deficiency with null space equal to a ``one-hot'' indicator vector, e.g. setting $\psi_3=0$ gives $\text{Null}(\Delta_{\text{Shapiro}}) = [\begin{array}{cccccc}
0&0&0&0&0 &1
\end{array}]^T$. However, \citet{shapiro_investigation_1983} pointed out that even with $\psi_3 = 0$, the system of equations $\boldsymbol{\sigma}(\boldsymbol{\theta}) = \boldsymbol{\sigma}$ can still be solved. Thus, similarly to the unidimensional function $f(x) = \text{sign}(x)\cdot x^3$ shown in  Figure \ref{fig:fx} having a unique maximum,  all parameters of the model, including $\psi_3$, are identifiable even though $\text{rk}(\Delta)<p$.  This is possible because both functions (the example function in Figure \ref{fig:fx} and the fitting function for SEM) have a discontinuous second derivative (Hessian) at the optimum.

While the above may seem like happy news, underidentification is not the only problem that can be caused by rank deficiencies. This can be seen by generating data from the model given above with the rank deficiency and fitting the model with free parameters to these data. For illustration purposes, I used 
 $\lambda_1 = 1$, $\lambda_1 = 0.4$, $\lambda_3 = 0.7$, 
 $\psi_1 = 1$, $\psi_2 = 0.3$, and $\psi_3 = 0$ as parameter settings and generated 2000 datasets with 100,000 cases each. 
 
\begin{figure}[tb]
    \centering
    \includegraphics{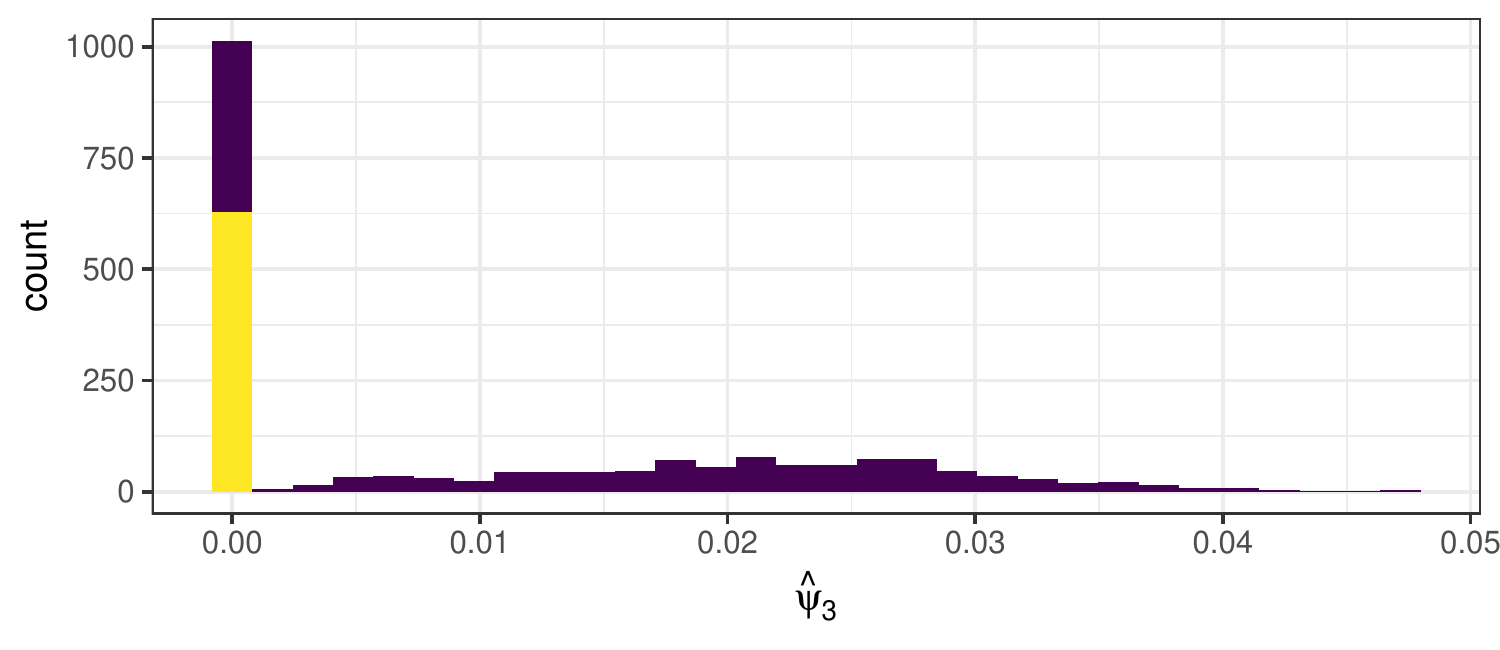}
    \includegraphics{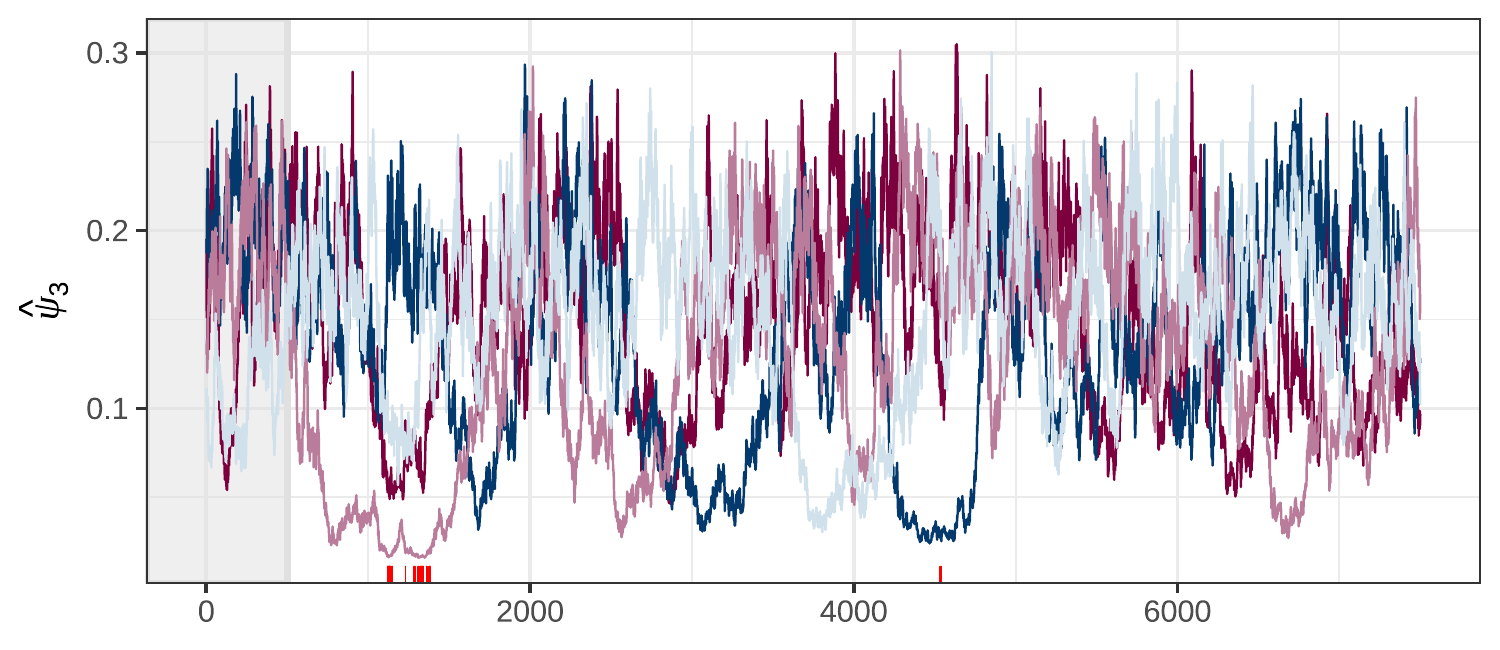}
    \caption{Empirical estimates of the parameter involved in Shapiro's rank deficiency at large samples. Yellow (light color) indicates ``inadmissible'' estimates ($\hat{\psi}_3 < 0$). Top: maximum-likelihood estimates. Bottom: a single Bayesian trace plot produced by \texttt{stan}. }
    \label{fig:shapiro}
\end{figure}

 The results are shown in Figure \ref{fig:shapiro} using maximum-likelihood (top panel) and a single run of Hamiltonian Monte Carlo using \texttt{stan} (bottom panel). In both cases the results lead to estimation problems. Using ML, the estimates $\hat{\psi}_3$ do not concentrate symmetrically around the true value, $\psi_3 = 0$. In spite of the very large number of cases, due to the rank deficiency, there is a nonzero probability that $\psi_3$ takes on an arbitrarily large value. For those solutions that are near the true value (spike near zero), normal-theory standard errors are arbitrarily large or cannot be calculated, so that the user cannot tell that the estimates were accurate. 
 
To illustrate that these problems are not specific to the estimation method, or to maximum-likelihood, I ran the same model in the Hamiltonian Monte Carlo (HMC) sampler \texttt{stan}, a popular package for Bayesian modeling. The HMC chains in the bottom panel of Figure \ref{fig:shapiro} show poor mixing and exhibit bias due to the constraint that estimates should be positive. In other words, the ``admissibility'' of solutions has been traded for bias. This demonstrates that Bayesian estimation can solve the problem of inadmissibility only at the cost of a bias that does not go to zero as the sample size grows. In addition, those draws that are actually near the rank-deficient true value are marked by \texttt{stan}'s HMC sampler as divergences.  Again, the user would be warned off these most accurate estimates.

These problems occur \emph{only} with the estimates  $\hat{\psi_3}$. Standard results, such as convergence to normality, large-sample unbiasedness, and correctness of normal-theory standard errors, do accrue for every other  parameter of the model. The reason for this is that the null space of $\Delta$, namely $\text{Null}(\Delta_{\text{Shapiro}}) = [\begin{array}{cccccc}
0&0&0&0&0 &1
\end{array}]^T$, does not involve any of these other parameters. We have noted a similar phenomenon in the CTUM SB-MTMM model, where the correlations among traits are not involved in any rank deficiency.

In short, although \citet{shapiro_investigation_1983} were correct in pointing out there is no \emph{identification} problem with this model, real-life analysis of this model using finite sample will yield plenty of other problems, regardless of the estimation method used. 
The same phenomenon occurs in more complex models such as the reduced-group split-ballot multitrait-multimethod model, discussed in the following section.

\subsection{Convergence and ``admissibility'' of SB-MTMM model}

The previous section illustrated the basic problems that result from  a simple rank deficiency in a simple model. 
I now illustrate how rank deficiencies affect estimation of the reduced-group split-ballot multitrait-multimethod model.

I generated data from a $3\times 3$ CTUM-MTMM model with equal loadings, and correlations that differed by a distance $\delta$:
\begin{align}
\Lambda = \kbordermatrix{
 &   \eta_1 &\eta_2 & \eta_3 & \xi_1& \xi_2& \xi_3\\
 y_{11} &    1 & 0 & 0  & 1 & 0 & 0  \\
 y_{12} &    1 & 0 & 0  & 0 & 1 & 0  \\
 y_{13} &     1 & 0 & 0  & 0 & 0 & 1  \\
 y_{21} &    0 & 1 & 0  & 1 & 0 & 0  \\
 y_{22} &     0 & 1 & 0  & 0 & 1 & 0  \\
 y_{23} &     0 & 1 & 0  & 0 & 0 & 1  \\
 y_{31} &     0 & 0 & 1  & 1 & 0 & 0  \\
 y_{32} &     0 & 0 & 1  & 0 & 1 & 0  \\
 y_{33} &     0 & 0 & 1  & 0 & 0 & 1}, & \; &
    \Phi =  \kbordermatrix{
     &   \eta_1 &\eta_2 & \eta_3 & \xi_1& \xi_2& \xi_3\\
 \eta_1  &  1 & 0.5 - \delta & 0.5  & 0& 0& 0\\
 \eta_2 &  0.5 - \delta & 1 & 0.5 + \delta & 0& 0& 0\\
 \eta_3  &  0.5 & 0.5 + \delta &1 & 0& 0& 0\\
  \xi_1 &  0& 0& 0 & 1& 0& 0\\
 \xi_2  &  0& 0& 0 & 0& 1& 0\\
  \xi_3 &  0& 0& 0 & 0& 0& 1\\
}, &\;&  \Psi = \mathbf{I}_9,
\end{align}
and data were generated from the standard confirmatory factor model, 
\begin{align}
    \mathbf{y} = \Lambda \boldsymbol{\eta} + \boldsymbol{\epsilon}, && \text{with} \; 
    \boldsymbol{\eta} \sim \text{MVN}(\mathbf{0}, \Phi) &&\text{and} \;
        \boldsymbol{\epsilon} \sim \text{MVN}(\mathbf{0}, \Psi).
\end{align}
When the SB-MTMM model is applied to this population, its Jacobian is rank deficient, $\text{rk}(\Delta) < p$, when $\delta = 0$.

Conditions were then defined by fully crossing the following two factors:
\begin{itemize}
    \item Sample size $n \in \{50, 75, 100, 500, 10^3, 10^4, 10^5\}$
    \item Distance from rank deficiency $\delta \in \{0, 0.01, 0.05, 0.1, 0.2, 0.3\}$
\end{itemize}
For each of the $7 \times 6 = 42$ conditions, $2000$ datasets were generated by sampling the nine observed variables jointly from a multivariate normal distribution. To simulate the planned missing data design, the first half of each dataset set all values for variables $y_{13}$, $y_{23}$,  and $y_{33}$ to ``missing''; the same was done for $y_{12}$, $y_{22}$,  and $y_{32}$ in the second half of each dataset. Using \texttt{lavaan} 0.6-5 \citep{rosseel_lavaan_2012}, I fit a CTUM MTMM model to each synthetic split-ballot dataset using maximum-likelihood under ignorability (``full-information maximum likelihood''). I then recorded whether the model converged using the default tolerance and whether the solution was ``admissible'' -- i.e. whether all variance matrices were positive-definite. \texttt{R} code for the simulation can be found in the Appendix. 

\begin{figure}[tb]
    \centering
    \includegraphics[width=\textwidth]{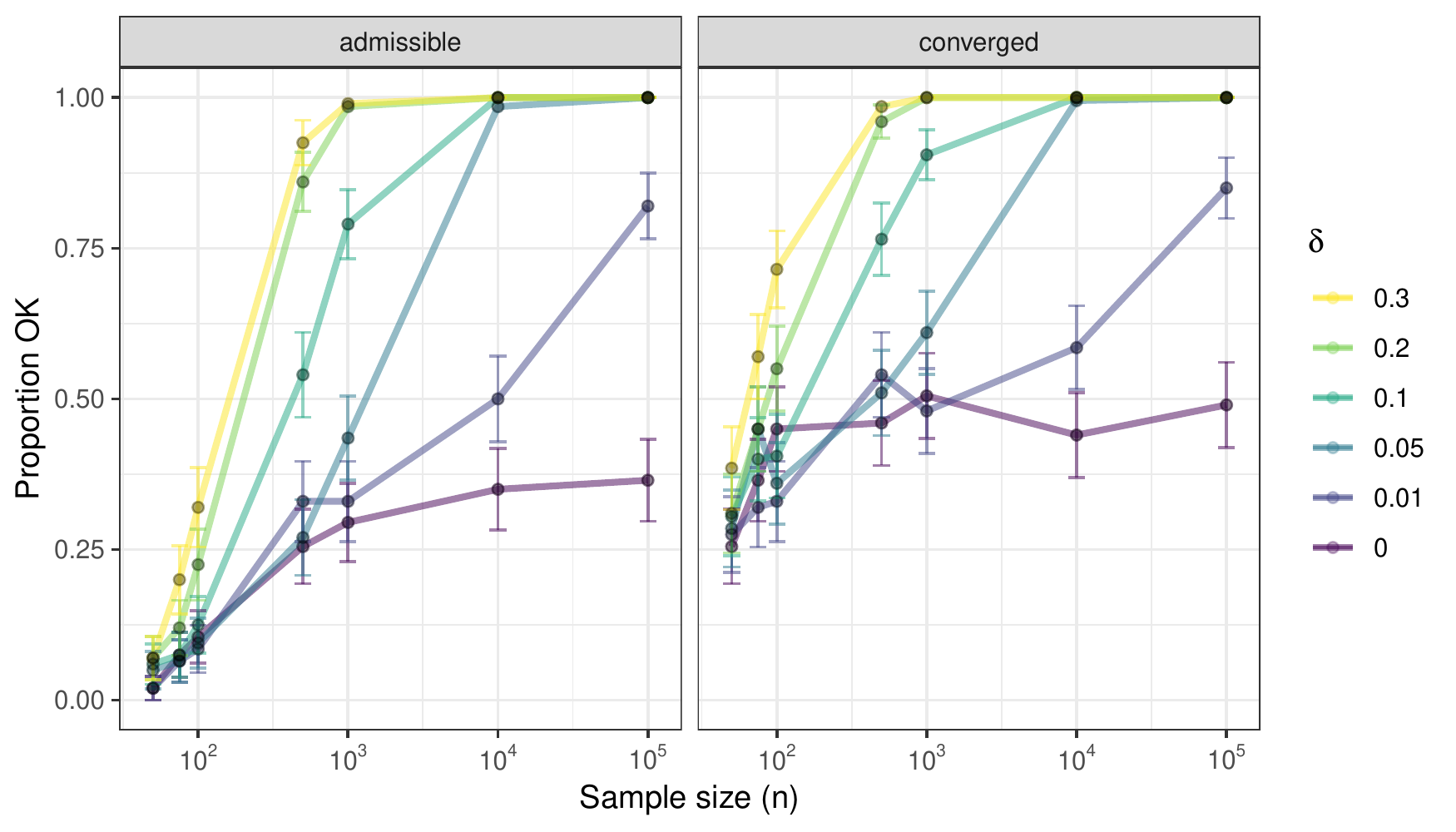}
    \caption{``Admissibility'' and convergence of the split-ballot CTUM MTMM model for different sample sizes and distances $\delta$ from the rank-deficient point.}
    \label{fig:results-simulation-mtmm}
\end{figure}

Figure \ref{fig:results-simulation-mtmm} shows the results of the simulation study. 

The left-hand side of Figure \ref{fig:results-simulation-mtmm} shows the proportion of "admissible" estimate sets. As expected, this proportion simply increases as the sampling variance decreases. As shown above, distance to the rank deficiency, $\delta$, interacts with the sample size to increase the admissible solutions; the further the population is from the rank deficiency, the stronger the influence of sample size. An exceptional case is the rank-deficient point itself, $\delta = 0$; at this point increasing the sample size does not ultimately lead to 100\% admissibility, regardless of sample size. For smaller distances, enormous sample sizes are needed to counteract the variance inflation. For example, at $\delta=0.01$, even 100,000 cases is not enough to yield 80\% admissible solution sets. 

The right-hand side of Figure \ref{fig:results-simulation-mtmm} shows the proportion of solutions that were deemed to have converged by the optimizer, \textsc{nl2sol} \citep{Dennis:1981:ANL:355958.355965}. Note that \textsc{nl2sol} employs several convergence tests simultaneously, including an explicit test for singularity of the Hessian. As for the admissibility, distance to the singular point $\delta$ interacts with the sample size. 
In addition, a paradoxical phenomenon can be observed at low values of $\delta$ and smaller sample sizes: there are points at which increasing the sample size \emph{decreases} the convergence. This happens because MLE's of parameters close to the rank deficient point can lie far away from this point when the sampling variance is large. In other words, the model converges more often because estimates may be far from the truth. 

\begin{figure}[tb] 
    \centering
    \includegraphics[width = 0.9\textwidth]{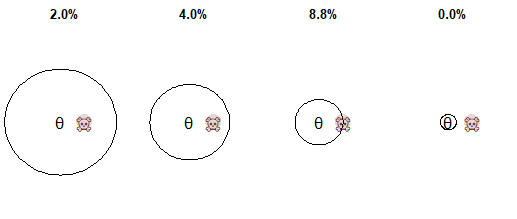}\\
    \includegraphics[width = 0.9\textwidth]{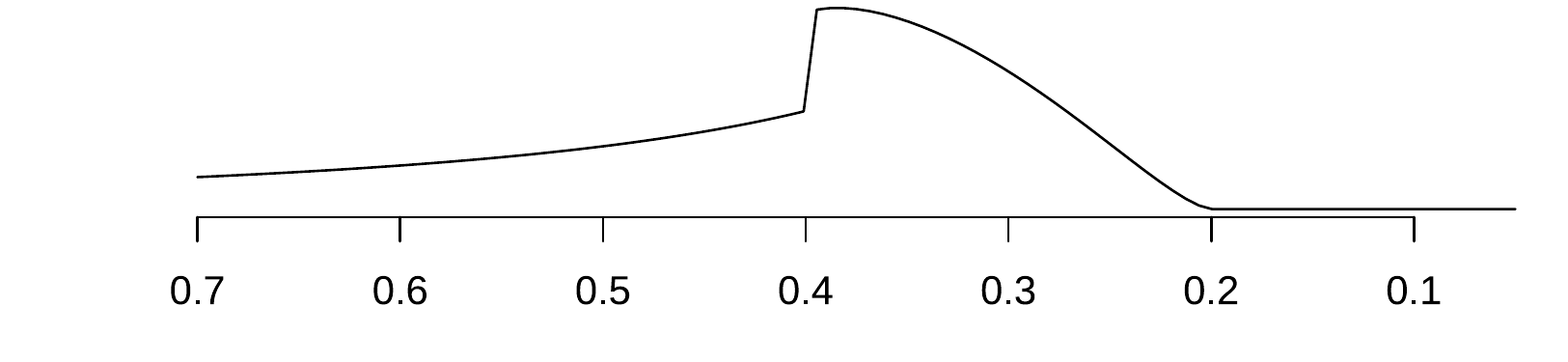}
    \caption{As the variance of the MLE decreases, the probability that it is located in an area that will lead to nonconvergence first increases, then decreases.}
    \label{fig:proportion_deathzone}
\end{figure}

 Figure \ref{fig:proportion_deathzone} demonstrates why nonmonotone effects can occur in the results in Figure \ref{fig:results-simulation-mtmm}. 
The Figure illustrates the overlap, in percentage area, between an arbitrary contour of the MLE (solid circle) around its true value (dark point) with an arbitrary region (red filled circle) leading to convergence problems around the rank deficient point (cross). The size of the contour (radius of the circle) depends on the variance of the MLE. The size of the region of nonconvergence will depend on the optimizer and choices regarding tolerance: lower tolerance will lead to larger red shaded regions. The bottom part of Figure \ref{fig:proportion_deathzone} plots the proportion overlap between two such areas as a function of the radius of the circle (MLE variance). It can be seen that this overlap shows the nonmonotone pattern found in the experiments. With large variance, fewer nonconvergence problems can occur, but only because the estimates are far from the true values. In these cases, ``inadmissible'' solutions (or bias when these are prevented using priors or restrictions) will also be more prevalent.

\section{Conclusion}

Even when a latent variable model is correctly specified, the model is identified, and the sample size is in the thousands or tens of thousands of cases, the sample may still very often  lack information about some of its parameters. In the literature on MTMM models, this phenomenon has often been noted under such descriptions as ``nonconvergence'', ``empirical underidentification'', ``inadmissible estimates'', and ``parameter instability''.  This paper has investigated deficient-rank Jacobians as the common cause for these ails.  Using the reduced-group split-ballot multitrait-multimethod model as a use case, we have demonstrated how rank deficiencies can be found and investigated using a computer algebra system. 

We found that the rank deficiency of the RG-SB-MTMM model does not extend to the estimation of the trait correlations. In other words, for all their problems, when the CTUM model is deemed acceptable by the researcher, and the trait correlations are the parameters of interest, the best course of action in the face of nonconvergence, inadmissible estimates, and their equivalent problems is to simply ignore them. Of course, this no longer holds when the (standardized) loadings or variance parameters themselves are of interest. Similarly, the result does not apply to CTCM models, which do generate a dependency involving the trait correlations. Nevertheless, it is clear that not all parameters are equally affected by estimation problems; an interesting conjecture is therefore that the trait correlations are less affected by these problems in general, even in CTCM or CTM models. 

The role of sample size was found to be more complex than previously thought. Due to the dual role of the rank deficiency, in increasing the variance and generating nonconvergence, increasing the sample size can sometimes deteriorate convergence. However, this does not mean that smaller sample sizes are better; rather, it means convergence is only achieved far away from the true value. Because convergence probability and the value of the parameter estimate are closely related, when running simulation studies, it should be common practice to report summaries for both the converged and the nonconverged estimates. Because previous simulation studies have reported bias, MSE, and variance for the converged estimates \emph{only}, they were unable to detect empirically that trait correlations were unbiased. 

Our results also suggest that estimation problems are fundamental properties of the model and available information in the data, and cannot be solved with computational ``tricks''. For instance, after estimating the Shapiro model using a modern HMC sampler, as commonly used in Bayesian modeling software, we found biased estimates and divergent transitions near the true value. A different potential solution suggested in recent literature is to employ Bayesian priors or penalized estimation  \citep[see][and references therein]{van_kesteren_structural_2019}. Employing such regularization methods can stabilize the estimates that are affected by the rank deficiency; however, one should be extremely careful to prevent the regularization from being applied to parameters that do not need it, such as the trait correlations in this example. We would therefore suggest that methodologists who wish to use regularization methods should consider these in the specific context of rank deficiencies in the Jacobian; ideally, the regularization should remove these, while not introducing bias in unaffected parameters. Similarly, the effect of alternative, potentially misspecified, model formulations, such as including covariates or removing one method or trait factor, should be carefully considered in this light.

MTMM is an old idea that has never been more relevant. As novel data sources flood into the social, behavioral, and biomedical sciences, it is more important than ever to evaluate the extent to which a combination of these different sources can provide us with valid and reliable measurement. To accomplish this goal, latent variable models, whether they be linear factor models or more modern (and complex) versions such as latent class MTMM \citep{oberski_latent_2015,oberski_evaluating_2017}, variational autoencoders, restricted Boltzman machines, or generative adversarial networks, are extremely useful tools. They also make demands on the data that can often not be met in practice. In the future, we hope that combining theory referred to in the present paper with practical goals of multi-source measurement will help overcome the barriers to leveraging the power of latent variables.

\bibliography{identification.bib}

\appendix

\section{Jacobian of the CTUM reduced-group split-ballot multitrait-multimethod model}

\begin{multline}
\label{eq:delta-ctum}
\boldsymbol{\Delta}_{\text{\sc{sbmtmm}}} =\\
\resizebox{1.1\hsize}{!}{
\kbordermatrix{  & \lambda_{11} & \lambda_{21} & \lambda_{31} & \lambda_{12} & \lambda_{22} & \lambda_{32} & \lambda_{13} & \lambda_{23} & \lambda_{33} & \psi_{1} & \psi_{2} & \psi_{3} & \psi_{4} & \psi_{5} & \psi_{6} & \psi_{7} & \psi_{8} & \psi_{9} & \rho_{12} & \rho_{13} & \rho_{23} & \phi_{4} & \phi_{5} & \phi_{6} \\
 \lambda_{11}^2+\psi_{1}+\phi_{4} & 2 \lambda_{11} & 0 & 0 & 0 & 0 & 0 & 0 & 0 & 0 & 1 & 0 & 0 & 0 & 0 & 0 & 0 & 0 & 0 & 0 & 0 & 0 & 1 & 0 & 0 \\
 \lambda_{11} \lambda_{12} & \lambda_{12} & 0 & 0 & \lambda_{11} & 0 & 0 & 0 & 0 & 0 & 0 & 0 & 0 & 0 & 0 & 0 & 0 & 0 & 0 & 0 & 0 & 0 & 0 & 0 & 0 \\
 \lambda_{11} \lambda_{21} \rho_{12}+\phi_{4} & \lambda_{21} \rho_{12} & \lambda_{11} \rho_{12} & 0 & 0 & 0 & 0 & 0 & 0 & 0 & 0 & 0 & 0 & 0 & 0 & 0 & 0 & 0 & 0 & \lambda_{11} \lambda_{21} & 0 & 0 & 1 & 0 & 0 \\
 \lambda_{11} \lambda_{22} \rho_{12} & \lambda_{22} \rho_{12} & 0 & 0 & 0 & \lambda_{11} \rho_{12} & 0 & 0 & 0 & 0 & 0 & 0 & 0 & 0 & 0 & 0 & 0 & 0 & 0 & \lambda_{11} \lambda_{22} & 0 & 0 & 0 & 0 & 0 \\
 \lambda_{11} \lambda_{31} \rho_{13}+\phi_{4} & \lambda_{31} \rho_{13} & 0 & \lambda_{11} \rho_{13} & 0 & 0 & 0 & 0 & 0 & 0 & 0 & 0 & 0 & 0 & 0 & 0 & 0 & 0 & 0 & 0 & \lambda_{11} \lambda_{31} & 0 & 1 & 0 & 0 \\
 \lambda_{11} \lambda_{32} \rho_{13} & \lambda_{32} \rho_{13} & 0 & 0 & 0 & 0 & \lambda_{11} \rho_{13} & 0 & 0 & 0 & 0 & 0 & 0 & 0 & 0 & 0 & 0 & 0 & 0 & 0 & \lambda_{11} \lambda_{32} & 0 & 0 & 0 & 0 \\
 \lambda_{12}^2+\psi_{2}+\phi_{5} & 0 & 0 & 0 & 2 \lambda_{12} & 0 & 0 & 0 & 0 & 0 & 0 & 1 & 0 & 0 & 0 & 0 & 0 & 0 & 0 & 0 & 0 & 0 & 0 & 1 & 0 \\
 \lambda_{12} \lambda_{21} \rho_{12} & 0 & \lambda_{12} \rho_{12} & 0 & \lambda_{21} \rho_{12} & 0 & 0 & 0 & 0 & 0 & 0 & 0 & 0 & 0 & 0 & 0 & 0 & 0 & 0 & \lambda_{12} \lambda_{21} & 0 & 0 & 0 & 0 & 0 \\
 \lambda_{12} \lambda_{22} \rho_{12}+\phi_{5} & 0 & 0 & 0 & \lambda_{22} \rho_{12} & \lambda_{12} \rho_{12} & 0 & 0 & 0 & 0 & 0 & 0 & 0 & 0 & 0 & 0 & 0 & 0 & 0 & \lambda_{12} \lambda_{22} & 0 & 0 & 0 & 1 & 0 \\
 \lambda_{12} \lambda_{31} \rho_{13} & 0 & 0 & \lambda_{12} \rho_{13} & \lambda_{31} \rho_{13} & 0 & 0 & 0 & 0 & 0 & 0 & 0 & 0 & 0 & 0 & 0 & 0 & 0 & 0 & 0 & \lambda_{12} \lambda_{31} & 0 & 0 & 0 & 0 \\
 \lambda_{12} \lambda_{32} \rho_{13}+\phi_{5} & 0 & 0 & 0 & \lambda_{32} \rho_{13} & 0 & \lambda_{12} \rho_{13} & 0 & 0 & 0 & 0 & 0 & 0 & 0 & 0 & 0 & 0 & 0 & 0 & 0 & \lambda_{12} \lambda_{32} & 0 & 0 & 1 & 0 \\
 \lambda_{21}^2+\psi_{3}+\phi_{4} & 0 & 2 \lambda_{21} & 0 & 0 & 0 & 0 & 0 & 0 & 0 & 0 & 0 & 1 & 0 & 0 & 0 & 0 & 0 & 0 & 0 & 0 & 0 & 1 & 0 & 0 \\
 \lambda_{21} \lambda_{22} & 0 & \lambda_{22} & 0 & 0 & \lambda_{21} & 0 & 0 & 0 & 0 & 0 & 0 & 0 & 0 & 0 & 0 & 0 & 0 & 0 & 0 & 0 & 0 & 0 & 0 & 0 \\
 \lambda_{21} \lambda_{31} \rho_{23}+\phi_{4} & 0 & \lambda_{31} \rho_{23} & \lambda_{21} \rho_{23} & 0 & 0 & 0 & 0 & 0 & 0 & 0 & 0 & 0 & 0 & 0 & 0 & 0 & 0 & 0 & 0 & 0 & \lambda_{21} \lambda_{31} & 1 & 0 & 0 \\
 \lambda_{21} \lambda_{32} \rho_{23} & 0 & \lambda_{32} \rho_{23} & 0 & 0 & 0 & \lambda_{21} \rho_{23} & 0 & 0 & 0 & 0 & 0 & 0 & 0 & 0 & 0 & 0 & 0 & 0 & 0 & 0 & \lambda_{21} \lambda_{32} & 0 & 0 & 0 \\
 \lambda_{22}^2+\psi_{4}+\phi_{5} & 0 & 0 & 0 & 0 & 2 \lambda_{22} & 0 & 0 & 0 & 0 & 0 & 0 & 0 & 1 & 0 & 0 & 0 & 0 & 0 & 0 & 0 & 0 & 0 & 1 & 0 \\
 \lambda_{22} \lambda_{31} \rho_{23} & 0 & 0 & \lambda_{22} \rho_{23} & 0 & \lambda_{31} \rho_{23} & 0 & 0 & 0 & 0 & 0 & 0 & 0 & 0 & 0 & 0 & 0 & 0 & 0 & 0 & 0 & \lambda_{22} \lambda_{31} & 0 & 0 & 0 \\
 \lambda_{22} \lambda_{32} \rho_{23}+\phi_{5} & 0 & 0 & 0 & 0 & \lambda_{32} \rho_{23} & \lambda_{22} \rho_{23} & 0 & 0 & 0 & 0 & 0 & 0 & 0 & 0 & 0 & 0 & 0 & 0 & 0 & 0 & \lambda_{22} \lambda_{32} & 0 & 1 & 0 \\
 \lambda_{31}^2+\psi_{5}+\phi_{4} & 0 & 0 & 2 \lambda_{31} & 0 & 0 & 0 & 0 & 0 & 0 & 0 & 0 & 0 & 0 & 1 & 0 & 0 & 0 & 0 & 0 & 0 & 0 & 1 & 0 & 0 \\
 \lambda_{31} \lambda_{32} & 0 & 0 & \lambda_{32} & 0 & 0 & \lambda_{31} & 0 & 0 & 0 & 0 & 0 & 0 & 0 & 0 & 0 & 0 & 0 & 0 & 0 & 0 & 0 & 0 & 0 & 0 \\
 \lambda_{32}^2+\psi_{6}+\phi_{5} & 0 & 0 & 0 & 0 & 0 & 2 \lambda_{32} & 0 & 0 & 0 & 0 & 0 & 0 & 0 & 0 & 1 & 0 & 0 & 0 & 0 & 0 & 0 & 0 & 1 & 0 \\
 \lambda_{11}^2+\psi_{1}+\phi_{4} & 2 \lambda_{11} & 0 & 0 & 0 & 0 & 0 & 0 & 0 & 0 & 1 & 0 & 0 & 0 & 0 & 0 & 0 & 0 & 0 & 0 & 0 & 0 & 1 & 0 & 0 \\
 \lambda_{11} \lambda_{13} & \lambda_{13} & 0 & 0 & 0 & 0 & 0 & \lambda_{11} & 0 & 0 & 0 & 0 & 0 & 0 & 0 & 0 & 0 & 0 & 0 & 0 & 0 & 0 & 0 & 0 & 0 \\
 \lambda_{11} \lambda_{21} \rho_{12}+\phi_{4} & \lambda_{21} \rho_{12} & \lambda_{11} \rho_{12} & 0 & 0 & 0 & 0 & 0 & 0 & 0 & 0 & 0 & 0 & 0 & 0 & 0 & 0 & 0 & 0 & \lambda_{11} \lambda_{21} & 0 & 0 & 1 & 0 & 0 \\
 \lambda_{11} \lambda_{23} \rho_{12} & \lambda_{23} \rho_{12} & 0 & 0 & 0 & 0 & 0 & 0 & \lambda_{11} \rho_{12} & 0 & 0 & 0 & 0 & 0 & 0 & 0 & 0 & 0 & 0 & \lambda_{11} \lambda_{23} & 0 & 0 & 0 & 0 & 0 \\
 \lambda_{11} \lambda_{31} \rho_{13}+\phi_{4} & \lambda_{31} \rho_{13} & 0 & \lambda_{11} \rho_{13} & 0 & 0 & 0 & 0 & 0 & 0 & 0 & 0 & 0 & 0 & 0 & 0 & 0 & 0 & 0 & 0 & \lambda_{11} \lambda_{31} & 0 & 1 & 0 & 0 \\
 \lambda_{11} \lambda_{33} \rho_{13} & \lambda_{33} \rho_{13} & 0 & 0 & 0 & 0 & 0 & 0 & 0 & \lambda_{11} \rho_{13} & 0 & 0 & 0 & 0 & 0 & 0 & 0 & 0 & 0 & 0 & \lambda_{11} \lambda_{33} & 0 & 0 & 0 & 0 \\
 \lambda_{13}^2+\psi_{2}+\phi_{6} & 0 & 0 & 0 & 0 & 0 & 0 & 2 \lambda_{13} & 0 & 0 & 0 & 1 & 0 & 0 & 0 & 0 & 0 & 0 & 0 & 0 & 0 & 0 & 0 & 0 & 1 \\
 \lambda_{13} \lambda_{21} \rho_{12} & 0 & \lambda_{13} \rho_{12} & 0 & 0 & 0 & 0 & \lambda_{21} \rho_{12} & 0 & 0 & 0 & 0 & 0 & 0 & 0 & 0 & 0 & 0 & 0 & \lambda_{13} \lambda_{21} & 0 & 0 & 0 & 0 & 0 \\
 \lambda_{13} \lambda_{23} \rho_{12}+\phi_{6} & 0 & 0 & 0 & 0 & 0 & 0 & \lambda_{23} \rho_{12} & \lambda_{13} \rho_{12} & 0 & 0 & 0 & 0 & 0 & 0 & 0 & 0 & 0 & 0 & \lambda_{13} \lambda_{23} & 0 & 0 & 0 & 0 & 1 \\
 \lambda_{13} \lambda_{31} \rho_{13} & 0 & 0 & \lambda_{13} \rho_{13} & 0 & 0 & 0 & \lambda_{31} \rho_{13} & 0 & 0 & 0 & 0 & 0 & 0 & 0 & 0 & 0 & 0 & 0 & 0 & \lambda_{13} \lambda_{31} & 0 & 0 & 0 & 0 \\
 \lambda_{13} \lambda_{33} \rho_{13}+\phi_{6} & 0 & 0 & 0 & 0 & 0 & 0 & \lambda_{33} \rho_{13} & 0 & \lambda_{13} \rho_{13} & 0 & 0 & 0 & 0 & 0 & 0 & 0 & 0 & 0 & 0 & \lambda_{13} \lambda_{33} & 0 & 0 & 0 & 1 \\
 \lambda_{21}^2+\psi_{3}+\phi_{4} & 0 & 2 \lambda_{21} & 0 & 0 & 0 & 0 & 0 & 0 & 0 & 0 & 0 & 1 & 0 & 0 & 0 & 0 & 0 & 0 & 0 & 0 & 0 & 1 & 0 & 0 \\
 \lambda_{21} \lambda_{23} & 0 & \lambda_{23} & 0 & 0 & 0 & 0 & 0 & \lambda_{21} & 0 & 0 & 0 & 0 & 0 & 0 & 0 & 0 & 0 & 0 & 0 & 0 & 0 & 0 & 0 & 0 \\
 \lambda_{21} \lambda_{31} \rho_{23}+\phi_{4} & 0 & \lambda_{31} \rho_{23} & \lambda_{21} \rho_{23} & 0 & 0 & 0 & 0 & 0 & 0 & 0 & 0 & 0 & 0 & 0 & 0 & 0 & 0 & 0 & 0 & 0 & \lambda_{21} \lambda_{31} & 1 & 0 & 0 \\
 \lambda_{21} \lambda_{33} \rho_{23} & 0 & \lambda_{33} \rho_{23} & 0 & 0 & 0 & 0 & 0 & 0 & \lambda_{21} \rho_{23} & 0 & 0 & 0 & 0 & 0 & 0 & 0 & 0 & 0 & 0 & 0 & \lambda_{21} \lambda_{33} & 0 & 0 & 0 \\
 \lambda_{23}^2+\psi_{7}+\phi_{6} & 0 & 0 & 0 & 0 & 0 & 0 & 0 & 2 \lambda_{23} & 0 & 0 & 0 & 0 & 0 & 0 & 0 & 1 & 0 & 0 & 0 & 0 & 0 & 0 & 0 & 1 \\
 \lambda_{23} \lambda_{31} \rho_{23} & 0 & 0 & \lambda_{23} \rho_{23} & 0 & 0 & 0 & 0 & \lambda_{31} \rho_{23} & 0 & 0 & 0 & 0 & 0 & 0 & 0 & 0 & 0 & 0 & 0 & 0 & \lambda_{23} \lambda_{31} & 0 & 0 & 0 \\
 \lambda_{23} \lambda_{33} \rho_{23}+\phi_{6} & 0 & 0 & 0 & 0 & 0 & 0 & 0 & \lambda_{33} \rho_{23} & \lambda_{23} \rho_{23} & 0 & 0 & 0 & 0 & 0 & 0 & 0 & 0 & 0 & 0 & 0 & \lambda_{23} \lambda_{33} & 0 & 0 & 1 \\
 \lambda_{31}^2+\psi_{8}+\phi_{4} & 0 & 0 & 2 \lambda_{31} & 0 & 0 & 0 & 0 & 0 & 0 & 0 & 0 & 0 & 0 & 0 & 0 & 0 & 1 & 0 & 0 & 0 & 0 & 1 & 0 & 0 \\
 \lambda_{31} \lambda_{33} & 0 & 0 & \lambda_{33} & 0 & 0 & 0 & 0 & 0 & \lambda_{31} & 0 & 0 & 0 & 0 & 0 & 0 & 0 & 0 & 0 & 0 & 0 & 0 & 0 & 0 & 0 \\
 \lambda_{33}^2+\psi_{9}+\phi_{6} & 0 & 0 & 0 & 0 & 0 & 0 & 0 & 0 & 2 \lambda_{33} & 0 & 0 & 0 & 0 & 0 & 0 & 0 & 0 & 1 & 0 & 0 & 0 & 0 & 0 & 1 \\
}}
\end{multline}

\section{Simulation \texttt{R} code}

\begin{footnotesize}
\begin{verbatim}
library(tidyverse)
library(lavaan)
library(viridis)


sim_data <- function(n, mod, split_ballot = TRUE, ...) {

    ydf <- simulateData(mod, sample.nobs = n, ...)

    if(split_ballot) {
      half <- floor(n/2)
      ydf[1:half, c(3,6,9)] <- NA
      ydf[(half+1):n, c(2,5,8)] <- NA
    }

    ydf
}

get_mod <- function(d) {
  paste0("
      T1 =~ y1 + y2 + y3
      T2 =~ y4 + y5 + y6
      T3 =~ y7 + y8 + y9
  
      M1 =~ 1*y1 + 1*y4 + 1*y7
      M2 =~ 1*y2 + 1*y5 + 1*y8
      M3 =~ 1*y3 + 1*y6 + 1*y9
  
      M1 ~~ 0*M2 + 0*M3 + 0*T1 + 0*T2 + 0*T3
      M2 ~~        0*M3 + 0*T1 + 0*T2 + 0*T3
      M3 ~~               0*T1 + 0*T2 + 0*T3
  
      T1 ~~ start(", 0.5 - d,")*T2 + start(", 0.5 + d,")*T3 + 1*T1
      T2 ~~ start(0.5)*T3 + 1*T2
      T3 ~~ 1*T3
  ")
}

any_variances_negative <- function(fit) {
  th <- coef(fit)
  any(th[grep("~~", names(th))] < 0)
}

runit <- function(n, mod) {
  
  ydf_sb <- sim_data(n, mod = mod, split_ballot = TRUE)
  
  fit <- lavaan(mod, data = ydf_sb, missing = "ml", int.ov.free = TRUE,
    auto.var = TRUE, auto.fix.first = FALSE)
  data.frame(converged = fit@Fit@converged,
             admissible = !any_variances_negative(fit), 
             rbind(coef(fit)))
  
}

runsim <- function(n, d, nsim) {
  purrr::map_df(1:nsim, ~runit(n, mod = get_mod(d = d))) %>% cbind(n = n, d = d)
}

set.seed(3452)
nsim <- 2e2
conditions <- expand.grid(n = c(50, 75, 1e2, 500, 1e3, 1e4, 1e5), 
                          d = c(0, 0.01, 0.05, 0.1, 0.2, 0.3))

res <- purrr::pmap_df(conditions, 
    ~ runsim(n = ..1, d = ..2, nsim = nsim))

res %>% 
  tidyr::gather("measure", "outcome", 1:2) %>% 
  mutate(d = as.factor(d)) %>%
  group_by(n, d, measure) %>% 
  summarize(prop_good = mean(outcome),
            se = sqrt((prop_good * (1 - prop_good))/nsim),
            lo = prop_good - 2*se, hi = prop_good + 2*se) %>% 
  ggplot(aes(n, prop_good, ymin = lo, ymax = hi, group = d, colour = d)) + ylim(0,1) + 
  geom_point(alpha = 0.5) + geom_line(lty = 2) + geom_errorbar(alpha = 0.5) + 
  facet_wrap(~measure) +  ggplot2::scale_x_log10() + theme_bw() + 
  viridis::scale_color_viridis(discrete = TRUE) + 
  geom_smooth(se = FALSE, lwd = 2)
\end{verbatim}
\end{footnotesize}

\end{document}